\documentclass[aps,letterpaper,superscriptaddress,pra,showpacs,amsmath,preprint,floats]{revtex4-1}

\usepackage[dvips]{graphicx}
\usepackage{amssymb,amsfonts,amsmath}
\usepackage{subfigure}
\usepackage{pdfpages}
\usepackage{braket}
\usepackage{bm}
\usepackage[utf8]{inputenc}
\usepackage[T1]{fontenc}
\usepackage{lmodern} % load a font with all the characters
\usepackage{amssymb,amsfonts,amsmath}
\usepackage{mathtools}
\usepackage{alltt}
\usepackage{mathrsfs}

\usepackage{chngcntr}

\begin{document}

\title{Effects of Non-idealities and Quantization of the Center of Mass Motion on Symmetric and Asymmetric Collective States in a Collective State Atomic Interferometer}

\author{Resham Sarkar}
\email{rsarkar@u.northwestern.edu}
\affiliation{Department of Physics and Astronomy, Northwestern University, 2145 Sheridan Road, Evanston, IL 60208, USA}

\author{May E. Kim}
\affiliation{Department of Physics and Astronomy, Northwestern University, 2145 Sheridan Road, Evanston, IL 60208, USA}

\author{Renpeng Fang}
\affiliation{Department of Physics and Astronomy, Northwestern University, 2145 Sheridan Road, Evanston, IL 60208, USA}

\author{Yanfei Tu}
\affiliation{Department of EECS, Northwestern University, 2145 Sheridan Road, Evanston, IL 60208, USA}

\author{Selim M. Shahriar}
\affiliation{Department of Physics and Astronomy, Northwestern University, 2145 Sheridan Road, Evanston, IL 60208, USA}
\affiliation{Department of EECS, Northwestern University, 2145 Sheridan Road, Evanston, IL 60208, USA}

\begin{abstract}
We investigate the behavior of an ensemble of $N$ non-interacting, identical atoms, excited by a laser. In general, the $i$-th atom sees a Rabi frequency $\Omega_i$, an initial position dependent laser phase $\phi_i$, and a motion induced Doppler shift of $\delta_i$. When $\Omega_i$ or $\delta_i$ is distinct for each atom, the system evolves into a superposition of $2^N$ intercoupled states, of which there are $N+1$ symmetric and $(2^N-(N+1))$ asymmetric collective states. For a collective state atomic interferometer (COSAIN) we recently proposed, it is important to understand the behavior of all the collective states under various conditions. In this paper, we show how to formulate the properties of these states under various non-idealities, and use this formulation to understand the dynamics thereof. We also consider the effect of treating the center of mass degree of freedom of the atoms quantum mechanically on the description of the collective states, illustrating that it is indeed possible to construct a generalized collective state, as needed for the COSAIN, when each atom is assumed to be in a localized wave packet. The analysis presented in this paper is important for understanding the dynamics of the COSAIN, and will help advance the analysis and optimization of spin squeezing in the presence of practically unavoidable non-idealities as well as in the domain where the center of mass motion of the atoms is quantized.
\end{abstract}


\maketitle

%% \linenumbers

%% main text
\section{Introduction}
\label{sec:intro}

Atom interferometry is emerging as a very important avenue of precision metrology. It has been used as a gyroscope~\cite{Borde,Kasevich,Gustavson} as well as an accelerometer~\cite{Geiger,Stern}. It has also been used for accurate measurements of gravity~\cite{Chu,Rosi}, gradients in gravity~\cite{Snadden}, as well as gravitational red-shift~\cite{Muller}. Other applications include measurement of fine structure constants with high precision~\cite{Biraben,Cadoret}, as well as the realization of a matter-wave clock~\cite{Lan}. The rotation sensitivity of an atom interferometric gyroscope (AIG) is due to the phase difference between two paths arising from the Sagnac effect~\cite{Sagnac,KasevichChu,SlowLight}. This phase difference is proportional to the area enclosed by the interferometer as well as the mass of each atom.

\begin{figure}[h]
\centering
\includegraphics[scale=0.25]{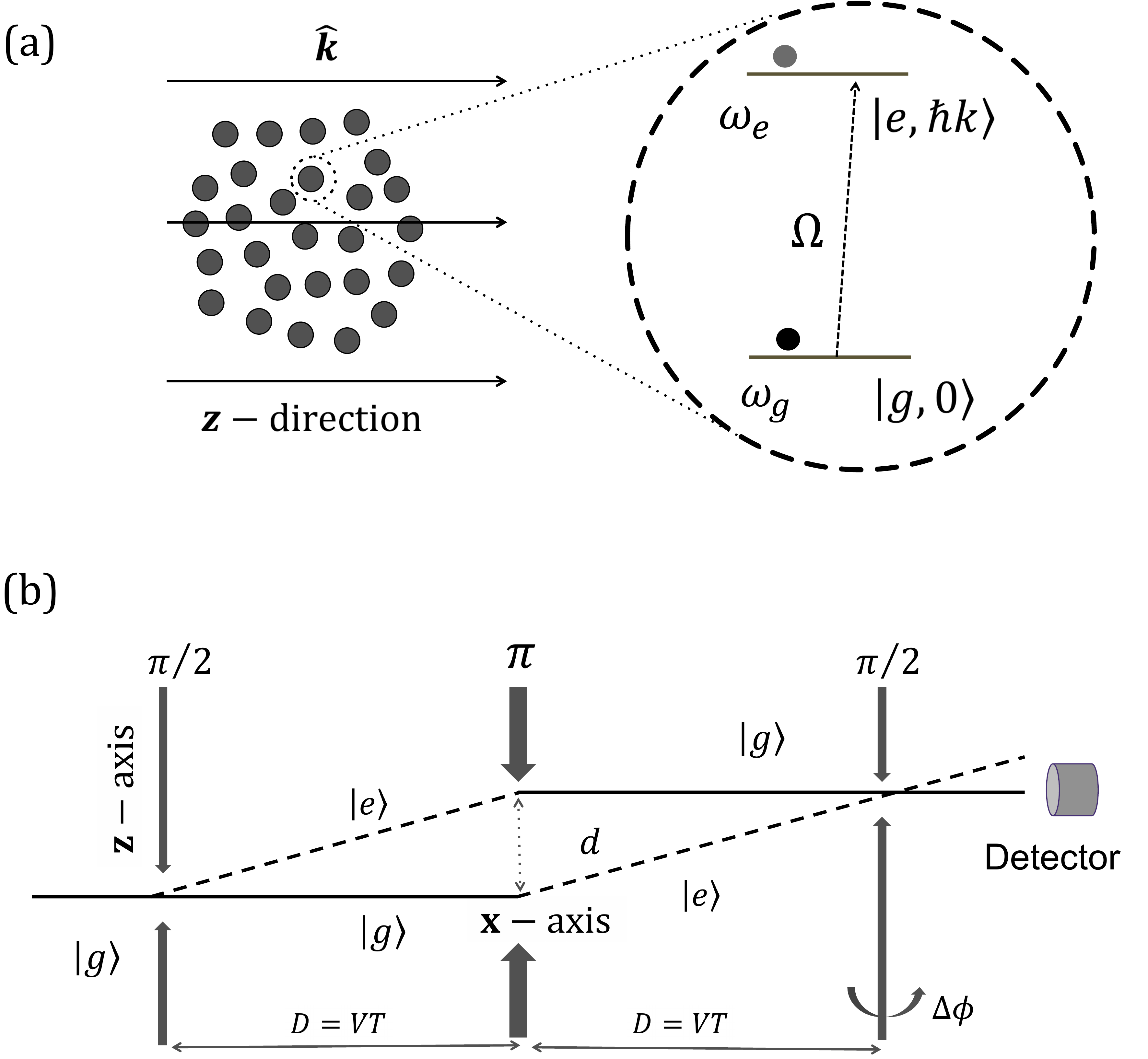}
\caption{\label{Fig_1}(a) Ensemble of $N$ two-level atoms in a classical laser field propagating the $\textbf{z}$-direction, (b) a single atom interferometer produced via $\pi/2-\pi-\pi/2$ sequence of excitation.}
\end{figure}

Motivated by this mass dependence of the rotation sensitivity of an AIG, we have recently proposed an interferometer that exploits the collective excitation of an ensemble of atoms~\cite{AIpaper}. To explain the principle behind this briefly, we consider an assembly of $N$ non-interacting identical two-level atoms, as illustrated in Fig.~\ref{Fig_1}(a). For a practical atomic interferometer, these levels are actually metastable hyperfine ground levels, coupled to an intermediate state via off-resonant counter-propagating optical fields. However, the basic concept can be illustrated by considering these two states to be coupled by a single, traveling laser field~\cite{BlochVector}. The atoms are initially prepared in quantum state $\ket{g,0}$, denoting that in this state the atoms are stationary along the $\textbf{z}$-axis. A laser beam propagating along the $\textbf{z}$-axis will impart a momentum $\hbar{k}$ to an atom upon absorption of a single photon, driving it to a superposition of the states $\ket{g,0}$ and $\ket{e,\hbar{k}}$, with the amplitude of each state depending on the intensity of the laser beam, $\Omega$, and the time of interaction, $t$. 

In a single atom interferometer, a two-level atom is first split into an equal superposition of $\ket{g,0}$ and $\ket{e,\hbar{k}}$ by a $\pi/2$-pulse (so that $\Omega{t}=\pi/2$). After letting this split atom drift freely for time $T$, the two states are inverted and redirected by a $\pi$-pulse. At the end of another free drift time, $T$, the two paths are recombined by another $\pi/2$-pulse. This is shown schematically in Fig.~\ref{Fig_1}(b). A possible phase difference, $\Delta\phi$ between the two paths manifests itself in the amplitude of the states at the end of the $\pi/2-\pi-\pi/2$ sequence. For example, the amplitude of $\ket{g}$ at the end of the interferometric sequence varies as $\cos^{2}(\Delta\phi/2)$~\cite{KasevichChu,Borde}. It is also possible to make a similar interferometer using only a single zone excitation~\cite{Shahriar,ShahriarExpt}

We have shown in ref.~\citenum{AIpaper} how an ensemble of $N$ atoms can be used to make a Collective State Atomic Interferometer (COSAIN) which also makes use of the $\pi/2-\pi-\pi/2$ pulse sequence employing counter propagating Raman excitations in a $\Lambda$ system, but has properties that differ very significantly from the Conventional Raman Atomic Interferometer (CRAIN) employing the same pulse sequence. For example, the width of the fringes generated as a function of the differential phase between the two paths (or, equivalently, a rotation applied perpendicular to its plane) is reduced by a factor of $\sqrt{N}$, when compared to the same for the CRAIN. The minimum measurable phase shift, under quantum noise limited (QNL) operation, is given by $\Delta\phi_c^{QNL}=\pi/\sqrt{Nn\tau\eta_c}$, where $n$ is the number of interrogations per unit time, $\tau$ is the total observation time, and $\eta_c$ is the quantum efficiency of detecting one of the collective states. This is to be compared with the same for a CRAIN, which is given by $\Delta\phi_s^{QNL}=\pi/\sqrt{m\tau\eta_s}$ where $m$ is the flux of atoms per unit time, and $\eta_s$ is the quantum efficiency of detecting each atom. For comparison, we consider a situation where $m = Nn$. Thus, $\Delta\phi_c^{QNL}$ can be substantially smaller than $\Delta\phi_s^{QNL}$, since $\eta_c$ can be very close to unity, while $\eta_s$ is generally very small because of geometric constraints encountered in collecting fluorescence from the atoms ~\cite{AIpaper}. In order to understand the basic principles of operation of such an interferometer, it is instructive to recall first the Dicke states~\cite{Dicke,Arecchi,Scully}.

\begin{figure}[h]
\centering
\includegraphics[scale=0.25]{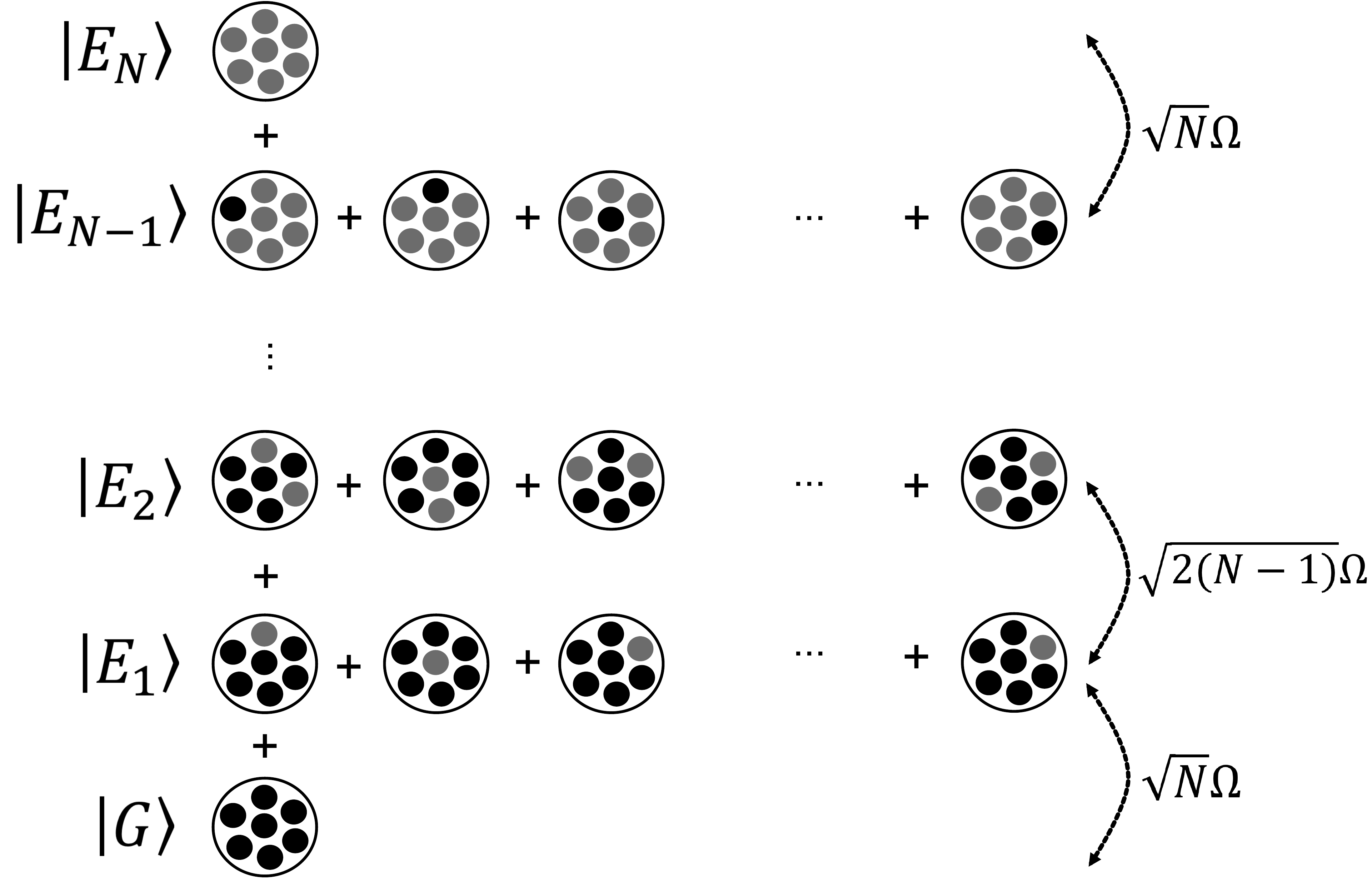}
\caption{\label{Fig_2} Schematic illustration of some of the possible symmetric collective states and coupling strength to their adjacent states.}
\end{figure}

In ref.~\citenum{Dicke}, Dicke showed that for a dilute ensemble of N atoms where the atoms do not interact, the ensemble evolves to a superposition of $N+1$ symmetric states (shown in Fig.~\ref{Fig_2}). Some of the possible Dicke states are defined as follows
\begin{align}
\centering
\ket{G} &= \ket{g_{1},g_{2},...,g_{N}},\nonumber\\
\ket{E_1} &={\sum}_{i=1}^{N}\ket{g_{1},g_{2},...,e_{i},...,g_{N}}\slash\sqrt{N},\nonumber\\
\ket{E_2} &= {\sum}_{j,k(j\not=k)}^{^{N \choose 2}}\ket{g_{1},...,e_{j},...e_{k},...,g_{N}}\slash\sqrt{{N \choose 2}},\nonumber\\
\ket{E_{N-1}} &= {\sum}_{i=1}^{N}\ket{e_{1},e_{2},...,g_{i},...,e_{N}}\slash\sqrt{N},\nonumber\\
\ket{E_N} &= \ket{e_{1},e_{2},...,e_{N}},
\label{Eqn_Collective}
\end{align}
etc. where ${N \choose n} = N!/n!(N-n)!$. 

A COSAIN is configured essentially the same way as the CRAIN, with two exceptions. First, it must make use of trapped atoms, released sequentially to the interferometer. Second, the detection process is designed to measure the probability of finding all the atoms in one of the collective states, such as $\ket{G}$. The reduction in the width of the fringe occurs due to a combination of the interferences among all the collective states, which follow different paths. Details of this process can be found in ref.~\citenum{AIpaper}. We have also shown how a Collective State Atomic Clock (COSAC) can be realized in this way, also with a $\sqrt{N}$ reduction in the width of the fringes~\cite{Clockpaper}. Just as the COSAIN is a variant of the CRAIN, the COSAC is a variant of the Conventional Raman Atomic Clock (CORAC)~\cite{Hemmer,Ezekiel,CsClock,Pati}. In ref.~\citenum{Clockpaper}, we show how a conventional microwave clock~\cite{Ramsey,Gibble} can also be converted into a COSAC.

As noted above, the COSAIN makes use of counter-propagating Raman transitions. As such, the characteristic wave number is $~k$, where $k=(k_1+k_2)$, and $k_1$ and $k_2$ are the wave numbers of the two laser beams. The non-zero temperature of a MOT provides a spread in the velocity of the constituent atoms. Therefore, each atom in the ensemble experiences a Doppler shift leading to a spread in detuning, with a zero mean value. Due to the finite size of the ensemble, each atom may experience a slightly different Rabi frequency depending on the spatial variation in the intensity profile of the laser beam. These factors contribute to a complex picture of an ensemble in a practical experiment. Furthermore, a semiclassical treatment of a quantum mechanical problem is not adequate. The wavepacket nature of the atoms must also be taken into account by considering the center of mass (COM) momentum of the atomic states.

In this paper, we present a description of collective states under generalized and non-ideal conditions, including a situation where the motion of the center of mass (COM) of each atom is treated quantum mechanically. Such a comprehensive model of the collective states currently does not exist in literature, and is important for understanding the behavior of the COSAIN. This comprehensive model of collective states, including the case where the COM motion is quantized, is also likely to help advance the analysis and optimization of spin squeezing~\cite{Ueda,Polzik,Kuzmich,SpinSqueezeRev}, under non-idealities that are unavoidable in any practical scheme. The rest of the paper is organized as follows. In Sec.~\ref{sec:semiclass} we describe the semiclassical model of generalized collective excitation to lay down the mathematical framework on which our analysis is based. For the sake of simplicity and transparency, we introduce the concepts first with the example of a $2$-atom ensemble identical to the Dicke formalism of collective excitation. Next, we analyze how variable Rabi frequencies and atomic velocities affect this simple ensemble. In Sec.~\ref{sec:Natom}, this investigation is extended to a general $N$-atom ensemble. In particular, we show that under certain conditions, the generalized asymmetric states of an ensemble are not decoupled from the symmetric set. We develop the general method of finding the generalized collective symmetric and asymmetric states in an ensemble of arbitrary size. In Sec.\ref{sec:QuantizedCOM}, we consider the COM motion degree of freedom of the atoms and investigate the implications of the wavepacket nature of the atoms, and therefore, of the ensemble. 

\section{Semiclassical Model of Generalized Collective Excitation}
\label{sec:semiclass}

Without loss of generality, we consider a collection of $N$ two level atoms, released from a cold trap, excited by a laser field traveling in the \textbf{z} direction, assuming the field amplitude to be of Gaussian profile in \textbf{x} and \textbf{y} directions, and constant in the \textbf{z} direction. Each atom is modeled as having two energy levels, $\ket{g_i}$ and $\ket{e_i}$. As mentioned earlier, a $\Lambda$-type atomic system excited by a pair of optically off-resonant laser fields propagating in opposite directions can be modeled as an effective two level system of this type~\cite{Shahriar}, so that the decay rate of the $\ket{e_i}$ state can be set to zero. This effective two-level system is shown in Fig 1(a), where $\omega_0=(\omega_e-\omega_g)$ is the frequency of the laser field, assumed to be resonant for stationary atoms. Each atom, however, experiences a different Doppler shift due to the thermal motion of the atoms, and consequently, a different effective laser frequency, $\omega_{0i}$. The net consequence of this is that the $i$-th atom picks up a detuning of $\delta_i$ depending on its velocity. The Rabi frequency, $\Omega_i$ experienced by the $i$-th atom depends on its position.

 The laser field is assumed, arbitrarily, to be polarized in the \textbf{x} direction. In the laboratory frame, the electric field at any point $\mathbf{r}=x\mathbf{\hat{x}}+y\mathbf{\hat{y}}+z\mathbf{\hat{z}}$, defined arbitrarily with respect to an origin, can be expressed as $\mathbf{E_i}(\mathbf{r},t)=\mathbf{\hat{x}}E_0\exp[-(x^2+y^2)/2\sigma_L^2]\cos(\omega_0t-kz)$, where $\sigma_L$ represents the width of the laser beam in the transverse directions. Assume now that, at $t=0$, the $i$-th atom is positioned at $\mathbf{r}_{0i}=x_{0i}\mathbf{\hat{x}}+y_{0i}\mathbf{\hat{y}}+z_{0i}\mathbf{\hat{z}}$, and is moving at a velocity $\mathbf{v}_i=v_{xi}\mathbf{\hat{x}}+v_{yi}\mathbf{\hat{y}}+v_{zi}\mathbf{\hat{z}}$. We ignore for now any change to this velocity due to the interaction with the laser field. This issue will be addressed later when we consider the motion of the COM of the atom quantum mechanically. In the reference frame of this atom, which is defined by the vector $\mathbf{r}_{i}=\mathbf{r}_{0i}+\mathbf{v}_it$, the electric field can be expressed as $\mathbf{E_i}(\mathbf{r_i},t)=\mathbf{\hat{x}}E_0\exp[-((x_{0i}+v_{xi}t)^2+(y_{0i}+v_{yi}t)^2)/2\sigma_L^2]\cos[\omega_0t-k(z_{0i}+v_{zi}t)]$. The transverse motion of the atom will lead to a time dependent variation of the amplitude of the Rabi frequency. We assume that, for typical systems of interest, $|v_{xi}t\ll\sigma_L|$ and $|v_{yi}t\ll\sigma_L|$, so that this variation can be ignored. We can then write the field seen by the atom in its reference frame as $\mathbf{E_i}(\mathbf{r},t)=\mathbf{\hat{x}}E_0\exp[-(x_{0i}^2+y_{0i}^2)/2\sigma_L^2]\cos(\omega_{0i}t-\xi_i)$, where $\omega_{0i}=\omega_0-kv_{zi}$ is the Doppler shifted frequency seen by the atom, and $\xi_i=kz_{0i}$ is a reference phase relation, determined by the initial position of the atom, between the atom and the field for all values of $t$.
 
  In the electric dipole approximation, the Hamiltonian for the $i$-th atom can be written as $H_i={\mathbf{P}_i}^{2}\slash2m+ H_{0i}+q\mathbf{\rho_i.E_i}$, where $\mathbf{P}_i$ is the COM momentum in the z-direction, $H_{0i}$ is the internal energy of the atom, $\mathbf{\rho}_i$ is the position of the electron with respect to the nucleus, $q$ is the electronic charge, and $m$ is the mass of the atom. As mentioned above, we are treating the motion of the COM of the atom semiclassically, deferring the quantum mechanical model thereof to a later part of this paper. As such, the COM term in the Hamiltonian can be ignored. Upon making the rotating-wave approximation, $H_i$ can then be expressed in the bases of $\ket{g_i}$ and $\ket{e_i}$ as:
\begin{align}
 H_i/\hbar =& \omega_g\ket{g_i}\bra{g_i}+ \omega_e\ket{e_i}\bra{e_i}+ \Omega_i(\exp(i(\omega_{0i}t-\xi_i))\ket{g_i}\bra{e_i}+ h.c.)/2,
\label{eq_SingleHam}
\end{align}
where $\Omega_i\equiv\bra{g_i}(\mathbf{x}\cdot\mathbf{\rho_i})\ket{e_i}E_i/\hbar$ $=\bra{e_i}(\mathbf{x}\cdot\mathbf{\rho_i})\ket{g_i}E_i/\hbar$.

The state of this atom, $\ket{\psi_i}$ evolves according to the Schr\"odinger equation, $i\hbar\partial\ket{\psi_i}/\partial{t}=H_i\ket{\psi_i}$. We define a transformed state vector $\ket{\psi_i'}=Q_i\ket{\psi_i}$, where $Q_i$ is a unitary transformation, defined as
\begin{equation}
 Q_i= \sum_{j=1}^{2}\exp(i(a_{ij}t+b_{ij}))\ket{j}\bra{j},
      \label{eq_SingleQ}
\end{equation}
where $a_{ij}$ and $b_{ij}$ are arbitrary parameters. The Hamiltonian for this state vector is then $H_i'=Q_iH_iQ_i^{-1}-\hbar\dot{Q_i}Q_i^{-1}$, so that $i\hbar\partial\ket{\psi_i'}/\partial{t}=H_i'\ket{\psi_i'}$. To render $H_i'$ time independent, we set $a_{i1} = \omega_g$ and $a_{i2} = \omega_{0i}+ \omega_g$. Now, setting $b_{i1} = 0$, $b_{i2} = -\xi_i$ makes $H_i'$ independent of any phase factor as well. In this frame, the $Q$-transformed Hamiltonian thus becomes
\begin{equation}
 H_i'/\hbar = -\delta_i\ket{e_i'}\bra{e_i'}+\Omega_i(\ket{g_i'}\bra{e_i'}+ h.c.)/2.
      \label{eq_SingleHam_Qtransformed}
\end{equation}
The new basis vectors, $\ket{g_i'}$ and  $\ket{e_i'}$, are related to the original basis vectors as $\exp(-i\omega_gt)\ket{g_i}$ and $\exp(-i((\omega_e+\delta_i)t-\xi_i))\ket{e_i}$, respectively. Assuming that the $i$-th atom is initially in the state $c_{gi}(0)\ket{g_i'}+c_{ei}(0)\ket{e_i'}$, its quantum state can be written as 
\begin{align}
\ket{\psi_i'} =&  e^{i\delta_i t/2}((c_{gi}(0)\cos\left(\frac{\Omega_i't}{2}\right)-i\frac{c_{gi}(0)\delta_i+c_{ei}(0)\Omega_i}{\Omega_i'}\sin\left(\frac{\Omega_i't}{2}\right))\ket{g_i'}\nonumber \\
& +(-i\frac{c_{gi}(0)\Omega_i-c_{ei}(0)\delta_i}{\Omega_i'}\sin\left(\frac{\Omega_i't}{2}\right)+c_{ei}(0)\cos\left(\frac{\Omega_i't}{2}\right))\ket{e_i'}),
\label{eq_SA_evolution_general}
\end{align}
where $\Omega_i'  = \sqrt{\Omega_i^2+\delta_i^2}$ is the effective coupling frequency of this atom.

Since we assume no interaction among the atoms, the ensemble Hamiltonian is the sum of all the individual Hamiltonians corresponding to each atom in the ensemble, $H_C'=\Sigma_iH_i'$. The state of the ensemble, therefore, evolves according to the Schr\"odinger equation, $i\hbar\partial\ket{\Psi_C'}/\partial{t}=H_C'\ket{\Psi_C'}$. For illustrative purposes, as well as transparency, let us consider first the case of $N=2$. $H_C'$ can be expressed as $H_1'{\otimes}I_2'+I_1'{\otimes}H_2'$, where $I_i'$ is the identity operator in the basis of $\ket{g_i'}$ and $\ket{e_i'}$ for the $i$-th atom. For instance, $\bra{g_1'g_2'}H_C'\ket{g_1'e_2'}=\bra{g_1'}H_1'\ket{g_1'}\bra{g_2'}I_2'\ket{e_2'}+\bra{g_1'}I_1'\ket{g_1'}\bra{g_2'}H_2'\ket{e_2'}=\bra{g_2'}H_2'\ket{e_2'}=\hbar\Omega_2/2$. Using this process, we can now express $H_C'$ in the basis of product states of the two atoms,  $\ket{g_1'g_2'}$, $\ket{e_1'g_2'}$, $\ket{g_1'e_2'}$ and $\ket{e_1'e_2'}$ as 
\begin{align}
H_C'/\hbar =& -\delta_1\ket{e_1'g_2'}\bra{e_1'g_2'}-\delta_2\ket{g_1'e_2'}\bra{g_1'e_2'}-(\delta_1+\delta_2)\ket{e_1'e_2'}\bra{e_1'e_2'}\nonumber\\
&+\Omega_1(\ket{g_1'g_2'}\bra{e_1'g_2'}+\ket{e_1'e_2'}\bra{g_1'e_2'}+h.c.)/2+\Omega_2(\ket{g_1'g_2'}\bra{g_1'e_2'}\nonumber\\
&+\ket{e_1'e_2'}\bra{e_1'g_2'}+h.c.)/2.
\label{eq_Hamiltonian_TwoAtoms}
\end{align}

Consider first the case where all the Rabi frequencies are the same, and there are no detunings. The $Q$-transformed Hamiltonian for each atom is then formally identical, since the phase factors due to different positions are encoded in the transformed basis states $\ket{g_i'}$ and $\ket{e_i'}$. Thus, the coupled collective states would now be formally identical to the symmetric Dicke states. For example,
\begin{align}
\centering
\ket{G'} =& \ket{g_1'}\ket{g_2'}, \nonumber\\
\ket{E_1'} =& (\ket{g_1'e_2'}+\ket{e_1'g_2'})/\sqrt{2}, \nonumber\\
\ket{E_2'} =& \ket{e_1'}\ket{e_2'}.
\end{align}

\begin{figure}[h]
\centering
\includegraphics[scale=0.18]{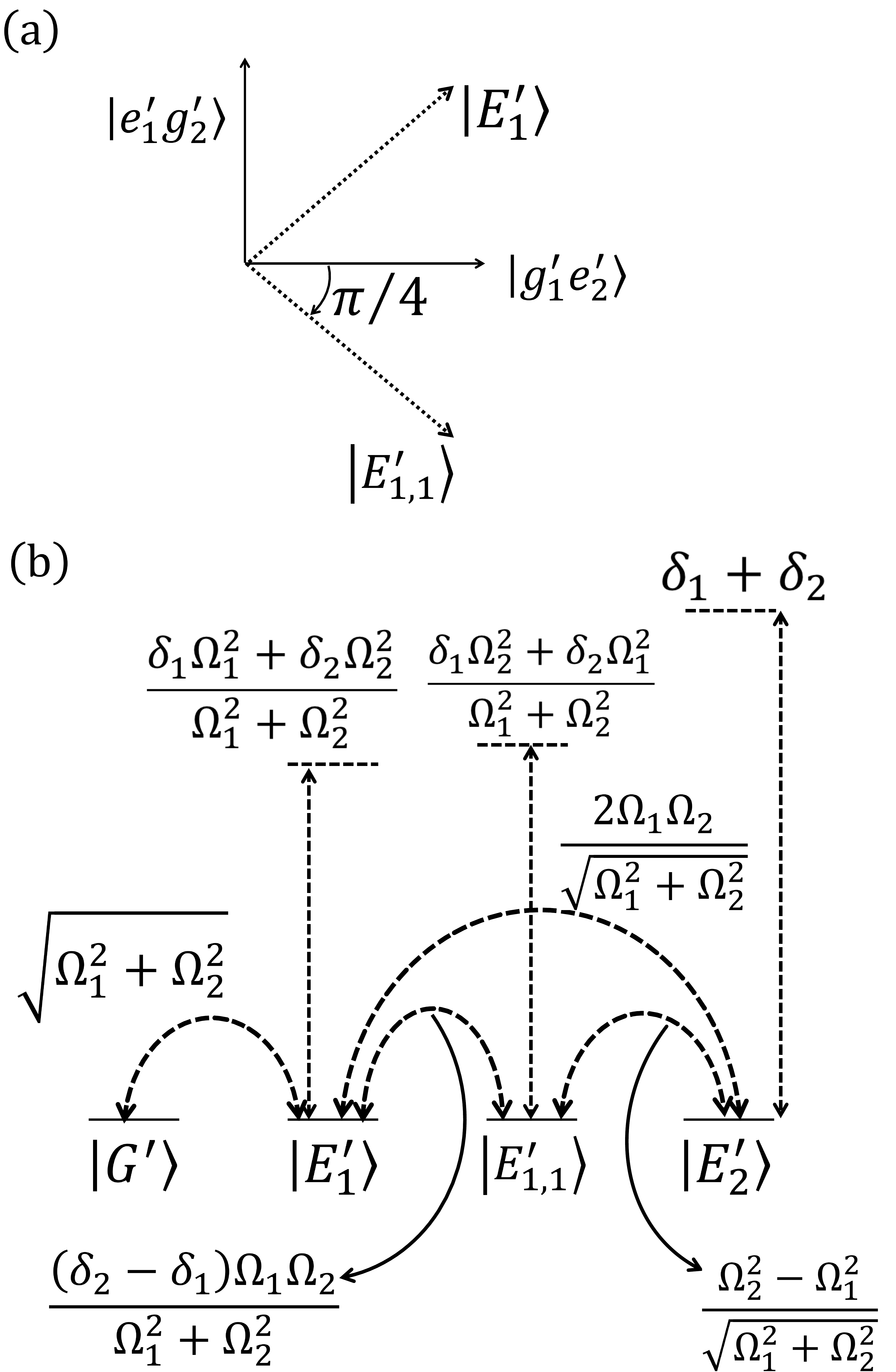}
\caption{\label{Fig_3} (a) Rotation of basis states to form collective states in a two-atom ensemble, (b) the complete set of all collective states and relevant couplings and detunings in a two-atom ensemble.}
\end{figure}

It should be noted that each of the constituent individual atomic states in these expressions include the temporal and spatial phase factors. Thus, these states behave the same way as the conventional Dicke  symmetric collective states, independent of the distance between the two atoms. It should also be noted that there exists another collective state, $\ket{E_{1,1}'}\equiv(\ket{g_1'e_2'}-\ket{e_1'g_2'})/\sqrt{2}$ which remains fully uncoupled from the symmetric set. The states $\ket{E_1'}$ and $\ket{E_{1,1}'}$ result from a $\pi/4$ rotation in the Hilbert subspace spanned by $\ket{e_1'g_2'}$ and $\ket{g_1'e_2'}$,  as illustrated in Fig. ~\ref{Fig_3}(a).

Consider next the case where there is still no detuning, but the Rabi frequencies are unequal. It is not obvious what the form of the symmetric collective states should be in this case. Consider first the task of finding the first excited symmetric collective state (SCS). Since the $\ket{G'}$ state will, by definition, be coupled only to this state, we can define this state, in general, as
\begin{equation}
\ket{E_1'}=\frac{H_C'\ket{G'}}{\sqrt{\braket{G' |H_C'^{\dagger}H_C'|G'}}},
\label{Eq_E1_from_G}
\end{equation}
where the denominator ensures that this state is normalized. When applied to the particular case at hand, we thus get $\ket{E_1'}=(\Omega_1\ket{e_1'g_2'}+\Omega_2\ket{g_1'e_2'})/\sqrt{\Omega_1^2+\Omega_2^2}$.

A rotation operator, $R$,  rotates the Hilbert sub-space, $\Phi_{2,1}$, formed by $\ket{e_1'g_2'}$ and $\ket{g_1'e_2'}$ by an angle $\theta=\tan^{-1}(\Omega_1/\Omega_2)$, such that one of the resulting states is $\ket{E_1'}$. This also produces a state $\ket{E_{1,1}'}=(\Omega_2\ket{e_1'g_2'}-\Omega_1\ket{g_1'e_2'})/\sqrt{\Omega_1^2+\Omega_2^2}$, which is orthogonal to $\ket{E_1'}$. In this rotated frame, the ensemble Hamiltonian, $\tilde{H_C'}=RH_C'R^{-1}$ becomes 
\begin{align}
\tilde{H_C'}/\hbar =& \sqrt{\Omega_1^2 +\Omega_2^2}\ket{G'}\bra{E_1'}/2+\Omega_1\Omega_2\ket{E_1'}\bra{E_2'}/\sqrt{\Omega_1^2+\Omega_2^2} \nonumber\\
&+(\Omega_2^2-\Omega_1^2)\ket{E_{1,1}'}\bra{E_2'}/2\sqrt{\Omega_1^2+\Omega_2^2} + h.c.
\label{eq_Hamiltonian_TwoAtoms_Collective}
\end{align}
Thus, the asymmetric collective state (ACS), $\ket{E_{1,1}'}$ does not remain isolated but is coupled to $\ket{E_2'}$, which, in turn is coupled to $\ket{E_1'}$. Consider next the case where we also allow for potentially different detunings for the two atoms, $\delta_1$ and $\delta_2$. It is easy to see, based on the general definition in Eq.~(\ref{Eq_E1_from_G}) of the first excited SCS, that $\ket{E_1'}$ has the same form as in Eq.~(\ref{Eq_E1_from_G}). Similarly, the expression for $\ket{E_{1,1}'}$ is also the same as above, and these states are generated by the same rotation operator, $R$, as given above. However, the coupling between the states in this rotated basis are now modified. Explicitly the ensemble $Q$-transformed Hamiltonian in the rotated frame becomes
\begin{align}
\tilde{H_C'}/\hbar =& -(\delta_1\Omega_1^2+\delta_2\Omega_2^2)(\ket{E_1'}\bra{E_1'}+\ket{E_{1,1}'}\bra{E_{1,1}'})/(\Omega_1^2+\Omega_2^2)\nonumber\\
&-(\delta_1+\delta_2)\ket{E_2'}\bra{E_2'}+\sqrt{\Omega_1^2+\Omega_2^2}\ket{G'}\bra{E_1'}/2\nonumber\\
&+\Omega_1\Omega_2\ket{E_1'}\bra{E_2'}/\sqrt{\Omega_1^2+\Omega_2^2} +(\Omega_2^2-\Omega_1^2)\ket{E_{1,1}'}\bra{E_2'}/2\sqrt{\Omega_1^2+\Omega_2^2}\nonumber\\
&- (\delta_1-\delta_2)\Omega_1\Omega_2\ket{E_1'}\bra{E_{1,1}'}/(\Omega_1^2+\Omega_2^2)+ h.c.
\label{Ham_Rotated_TwoAtoms_alldiff}
\end{align}
Thus, the ACS $\ket{E_{1,1}'}$ is now coupled directly to the SCS $\ket{E_1'}$, in addition to being coupled to the state $\ket{E_2'}$. Furthermore, the energies of the states are also shifted with respect to $\ket{G'}$. These couplings and shifts are illustrated in Fig. ~\ref{Fig_3}(b).

In an ensemble with a large number of atoms, the number of asymmetric states is far larger than that of the symmetric states. In the next section, we discuss a more generalized view of collective states, considering the variations in different parameters and manifestations thereof in the behavior of the collective states.

In the preceding discussions, we have taken into account the facts that each atom is at a unique position (which means that it sees a unique phase of the laser), sees a potentially unique Rabi frequency, and is moving with a particular velocity which in turn produces a Doppler shift. A natural question that may arise is whether we are taking into account the fact that the position of each atom is changing with time, so that it would see a time varying Rabi frequency and laser phase. The temporal variation in Rabi frequency can be ignored because the velocity of each atom is assumed to be very small. In appendix ~\ref{sec:append}, we show that the temporal change in the laser phase seen by the atom is akin to taking into account the Doppler shift.

\section{N-Atom Ensemble}
\label{sec:Natom}

The Hamiltonian of an ensemble of $N$ non-interacting and non-overlapping atoms is simply given by the sum of the Hamiltonians of the constituent atoms as noted above. It is convenient to express these as a sum of three parts: raising, lowering and detuning: $H_C'=H_{r}'+H_{l}'+H_{d}'$, where $H_{r}' = {\sum}_i^N\hbar\Omega_i\ket{e_i'}\bra{g_i'}/2$, $H_{l}' = {\sum}_i^N\hbar\Omega_i\ket{g_i'}\bra{e_i'}/2$, and $H_{d}' = -{\sum}_i^N\hbar\delta_i\ket{e_i'}\bra{e_i'}$. The raising part of the Hamiltonian, $H_{r}'$ couples $\ket{E_n'}$ to its adjacent higher SCS, $\ket{E_{n+1}'}$. Similarly, the lowering part of the Hamiltonian, $H_{l}'$ couples $\ket{E_n'}$ to its adjacent lower SCS, $\ket{E_{n-1}'}$. The function of the third term, $H_{d}'$ is two fold. First, it leads to a shift in the energy of the collective states (symmetric and asymmetric). Second, under certain conditions, it leads to a coupling between the SCS and all the ACS's, as well as among all the ACS's, within the same manifold (i.e., the set of collective states corresponding to the absorption of a given number of photons). Analogous to Eq.~(\ref{Eq_E1_from_G}), $\ket{E_{n+1}'}$ can be generated from $\ket{E_n'}$, for any value of $n$, using the following prescription
 \begin{align}
 \ket{E_{n+1}'}=\frac{H_{r}'\ket{E_n'}}{\sqrt{\braket{E_n'|H_{r}'^{\dagger}H_{r}'|E_n'}}}.
 \label{Eq_En+1_from_En}
 \end{align}
 
To illustrate the use of Eq.~(\ref{Eq_En+1_from_En}), we first consider the ideal case where each atom sees the same Rabi frequency, and experiences no Doppler shift, but still allowing for the fact that different atoms see different spatial phases. Since $H_d'=0$, the asymmetric states remain fully uncoupled from the symmetric states. Using Eq.~(\ref{Eq_En+1_from_En}), we can now easily find $\ket{E_1'}$, noting that $\ket{E_0'}=\ket{G'}$. Application of $H_r'$ to $\ket{G'}$, upon normalization, then leads to the result that $\ket{E_1'}={\sum}_{k=1}^N\ket{g_1'g_2',...e_k',...,g_N'}/\sqrt{N}$. This is essentially the same as the well-known first-excitation Dicke state, with the exception that the spatial phases seen by the individual atoms are incorporated in the constituent states $\ket{g'}$ and $\ket{e'}$, as noted before in the context of $N=2$.

It is now easy to see how to generate $\ket{E_n'}$ for any value of $n$, by repeated application of $H_r'$, and allowing for the normalization, as prescribed by Eq.~(\ref{Eq_En+1_from_En}). Specifically, we get
\begin{align}
\ket{E_n'}=J(N,n)^{-1/2}\sum_{k=1}^{J(N,n)}P_k\ket{g'^{\otimes(N-n)}e'^{\otimes n}},
\end{align}
where $J(N,n)\equiv J={N \choose n}$, and $P_k$ is the permutation operator \cite{HumeWineland}. 

Under the ideal condition being considered here, the ACS's remain fully decoupled from the symmetric set at all times, as noted above. As such, an explicit description of the forms of the ACS's is not necessary for understanding the behavior of the ensemble. However, when we consider non-idealities later, it will be important to understand the form of the ACS's. Therefore, we discuss here how to determine these states explicitly in the ideal case, and a simple modification of this approach will then be used later on for the non-ideal cases, where the ACS's are relevant.

Consider a particular manifold corresponding to the absorption of $n$ photons. The SCS is $\ket{E_n'}$, and there are $(J-1)$ ACS's, denoted as $\ket{E_{n,j}'}$ for $j=1$ to $(J-1)$. To find these states, we consider $\Phi_{N,n}$, the Hilbert sub-space of dimension $J$ spanned by the states $P_k\ket{g'^{\otimes(N-n)}e'^{\otimes n}}$. The elements of  $\Phi_{N,n}$ are arbitrarily labeled $\hat{s}_1,\hat{s}_2,\ldots,\hat{s}_J$. The SCS is a particular vector in this Hilbert space, and the ACS's are any set of mutually orthogonal vectors that are all normal to the SCS. Thus, the set of ACS's is not unique, and there are many ways to construct them. The standard procedure for finding such a set of orthonormal vectors is the Gram-Schmidt Orthogonalization (GSO) process. From a geometric point of view, the GSO process can be seen as a set of generalized rotations (with potentially complex angles) in the Hilbert space. Given that the SCS consists of a superposition of the basis vectors with real coefficients, these rotations can be viewed in terms of physical angles for $N=2$ and $3$, whereas for $N>3$, the angles have to be interpreted in an abstract manner. In order to elucidate our understanding of the ACS's, we first formulate the construction of ACS's for arbitrary $N$ and $n$, by  successive rotations of the Hilbert subspace, $\Phi_{N,n}$. We then illustrate the application of this model for $N=3$ for constructing some explicit version of the ACS's (noting that the $N=2$ case has only a single ACS which can be found trivially and has been explained in detail in Sec.~\ref{sec:semiclass}).

The elements of  $\Phi_{N,n}$, labeled $\hat{s}_1,\hat{s}_2,\ldots,\hat{s}_J$, form the coordinate axes of this Hilbert space. In this picture, we can represent the SCS as $\mathbf{V}=(\hat{s}_1+\hat{s}_2+\ldots+\hat{s}_J)/\sqrt{J}$, a vector that makes an angle, $\theta=\cos^{-1}(1/\sqrt{J})$ with each of the axes. Thus, to find all the collective states of $\Phi_{N,n}$, including the SCS and all the ACS's, we proceed as follows. We start with the original set of coordinate axes: $\hat{s}_1,\hat{s}_2,\ldots,\hat{s}_J$. We then carry out a set of $(J-1)$ rotations, producing a new set of coordinate axes that are mutually orthogonal. The rotation angles are chosen to ensure that after the $(J-1)$ rotations, one of the coordinate axes is parallel to $\mathbf{V}$ (which is the SCS), so that the remaining set of coordinate axes can be identified as the ACS's. This is accomplished by carrying out the following steps:

\textbf{Step 1:} We write $\mathbf{V}$ as a sum of two terms, $\mathbf{V}_{12}$ and $\mathbf{V}_{rest}$, where $\mathbf{V}_{12}=(\hat{s}_1+\hat{s}_2)/\sqrt{J}$. Normalization of $\mathbf{V}_{12}$ gives the unit vector $\mathbf{\hat{V}}_{12}=(\hat{s}_1+\hat{s}_2)/\sqrt{2}$, revealing that it makes an angle $\cos^{-1}(1/\sqrt{2})$ with $\hat{s}_1$ and $\hat{s}_2$. Therefore, the plane of $\hat{s}_1$ and $\hat{s}_2$ must be rotated around the origin by $\theta_2=(-\cos^{-1}(1/\sqrt{2}))$ to give $\hat{s}'_1=(\hat{s}_1-\hat{s}_2)/\sqrt{2}$ and $\hat{s}'_2=(\hat{s}_1+\hat{s}_2)/\sqrt{2}$. Obviously, $\hat{s}'_2$ is parallel to $\mathbf{V}_{12}$. By construction, $\hat{s}'_1$ is orthogonal to $\hat{s}'_2$, and therefore to $\mathbf{V}_{12}$. Since $\mathbf{V}_{rest}$ does not contain any component in the $\{\hat{s}_1,\hat{s}_2\}$ plane, it then follows that $\hat{s}'_1$ is orthogonal to $\mathbf{V}$, and is therefore an ACS. For $N=2$ described in Sec.~\ref{sec:semiclass}, $\hat{s}'_1=\ket{E_{1,1}'}$ and $\hat{s}'_2=\ket{E_1'}$, and the process stops at this point.

\begin{figure}[h]
\centering
\includegraphics[scale=0.20]{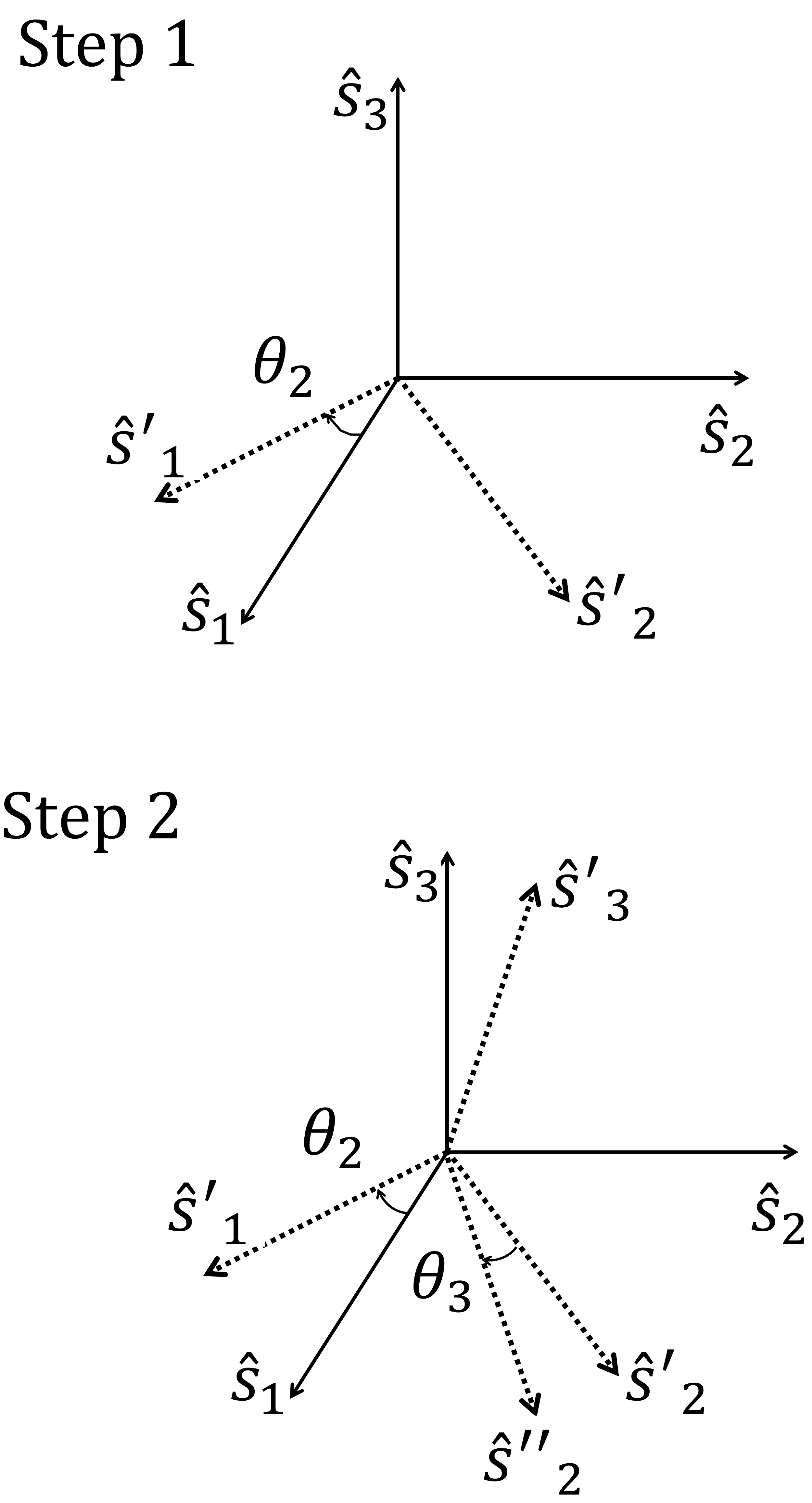}
\caption{\label{Fig_4} Hilbert sub-space rotation of the first excited state of an ensemble of three atoms.}
\end{figure}

\textbf{Step 2:} The vector, $\mathbf{V}$ is rewritten as another sum of two terms, $\mathbf{V}_{123}$ and $\mathbf{V}'_{rest}$, where $\mathbf{V}_{123}=(\hat{s}_1+\hat{s}_2+\hat{s}_3)/\sqrt{J}$. Normalization of $\mathbf{V}_{123}$ gives the unit vector $\mathbf{\hat{V}}_{123}=(\hat{s}_1+\hat{s}_2+\hat{s}_3)/\sqrt{3}$, showing that it makes an angle $\cos^{-1}(1/\sqrt{3})$ with $\hat{s}_1$, $\hat{s}_2$ and $\hat{s}_3$. Since $\hat{s}'_1$ is orthogonal to $\mathbf{V}$, we leave it undisturbed. The plane of $\hat{s}'_2$ and $\hat{s}_3$ is rotated around the origin by $\theta_3=(-\cos^{-1}(1/\sqrt{3}))$, resulting in $\hat{s}''_2=(\hat{s}_1+\hat{s}_2-2\hat{s}_3)/\sqrt{6}$ and $\hat{s'}_3=(\hat{s}_1+\hat{s}_2+\hat{s}_3)/\sqrt{3}$. It is clear that $\hat{s}'_3$ is parallel to $\mathbf{V}_{123}$. By construction,  $\hat{s}''_2$ is orthogonal to  $\hat{s}'_3$, and therefore, to $\mathbf{V}_{123}$. Furthermore, since $\mathbf{V}'_{rest}$ does not contain any component in the $\{\hat{s}_1,\hat{s}_2,\hat{s}_3\}$ plane, it then follows that $\hat{s}''_2$ is orthogonal to $\mathbf{V}$. $\hat{s}''_2$ is also orthogonal to $\hat{s}'_1$, since it is a linear combination of $\hat{s}'_2$ and $\hat{s}'_3$, which are both orthogonal to $\hat{s}'_1$. Thus, $\hat{s}''_2$ is the second ACS. For $N=3$ and $n=2$, this is the terminal step, resulting in $\hat{s}'_1=\ket{E_{2,1}'}$, $\hat{s}''_2=\ket{E_{2,2}'}$ and $\hat{s}'_3=\ket{E_2'}$, as shown in Fig.~\ref{Fig_4}. 

\textbf{Step 3:} $\mathbf{V}$ is written again as $\mathbf{V}=\mathbf{V}_{1234}+\mathbf{V}''_{rest}$, where $\mathbf{V}_{1234}=(\hat{s}_1+\hat{s}_2+\hat{s}_3+\hat{s}_4)/\sqrt{J}$. Again, normalizing $\mathbf{V}_{1234}$ gives $\mathbf{\hat{V}}_{1234}=(\hat{s}_1+\hat{s}_2+\hat{s}_3+\hat{s}_4)/\sqrt{4}$, showing that it makes an angle $\cos^{-1}(1/\sqrt{4})$ with $\hat{s}_1$, $\hat{s}_2$, $\hat{s}_3$ and $\hat{s}_4$. As described in Step 2 above, $\hat{s}'_1$ and $\hat{s}''_2$ are orthogonal to each other and to $\mathbf{V}$, and, therefore, we leave these two undisturbed. To find the vector orthogonal to this pair as well as to $\mathbf{V}$, we rotate the plane of $\hat{s}'_3$ and $\hat{s}_4$ about the origin by $\theta_4=(-\cos^{-1}(1/\sqrt{4}))$, and derive $\hat{s}''_3=(\hat{s}_1+\hat{s}_2+\hat{s}_3-3\hat{s}_4)/\sqrt{12}$ and $\hat{s}'_4=(\hat{s}_1+\hat{s}_2+\hat{s}_3+\hat{s}_4)/\sqrt{4}$. Following the same set of arguments presented in Step 2, it is easy to show that $\hat{s}''_3$ is orthogonal to $\hat{s}'_1, \hat{s}''_2$ and $\mathbf{V}$. As such, this is the third ACS. For $N=4$ and $n=1$, this is the terminal step, resulting in $\hat{s}'_1=\ket{E_{1,1}'}$, $\hat{s}''_2=\ket{E_{1,2}'}$, $\hat{s}''_3=\ket{E_{1,3}'}$ and $\hat{s}'_4=\ket{E_1'}$. 

For arbitrary $N$ and $n$, there are $(J-1)$ such steps to arrive at the Hilbert sub-space $\Phi_{N,n}'$ spanned by $\hat{s}'_1, \hat{s}''_2,\hat{s}''_3, \ldots, \hat{s}'_J$, where $\hat{s}'_J$ is the SCS and the rest are the ACS's. This process can be formalized by the method of matrix rotations considering the column vector formed by the elements of the space $\Phi_{N,n}$ as follows
\begin{align}
\mathbf{S}=\begin{bmatrix}\hat{s}_1& \hat{s}_2 \ldots & \hat{s}_J\end{bmatrix}^T.
\end{align}
The vector, $\mathbf{S}$ undergoes a series of rotations that transforms it to another vector, $\mathbf{S}_{C}$ whose elements are the symmetric and asymmetric collective states for that particular manifold of the ensemble. The first rotation matrix, $R(2)$ causes a rotation of $\mathbf{S}$ in the $\{\hat{s}_1,\hat{s}_2\}$ plane to form $\mathbf{S}_2$ whose elements are $\{\hat{s'}_1,\hat{s'}_2,\hat{s}_3,\ldots,\hat{s}_J\}$. The second rotation matrix, $R(3)$ further rotates the vector $\mathbf{S}_2$ in the $\{\hat{s'}_2,\hat{s}_3\}$ plane to give $\mathbf{S}_3$. This process is continued until the vector, $\mathbf{S}_J\equiv\mathbf{S}_{C}$ is formed by applying $R(J)$ on $\mathbf{S}_{J-1}$. Therefore, the overall process may be expressed as $\mathbf{S}_{C}=R_T\mathbf{S}$, where $R_T=R(J)R(J-1)\ldots R(3)R(2)$. The $j$-th rotation vector is of the form 
\begin{align}
R(j)_{m,n}&=\begin{cases}
 1\; \text{for} \; m=n, m\neq j-1,j\\
 \cos\theta_j \; \text{for} \; m = n = j-1, j \\
 -\sin\theta_j \; \text{for} \; m=j, n = j-1 \\
 \sin\theta_j \; \text{for} \; m=j-1, n = j \\
 0 \; \text{otherwise}
 \end{cases}, 
\end{align}
where $\theta_j = \cos^{-1}(1/\sqrt{j})$, so that $\cos\theta_j=1/\sqrt{j}$ and $\sin\theta_j=\sqrt{(j-1)/j}$. This matrix represents a simple rotation by an angle of $(-\theta_j)$ in the plane of $\hat{s}_{j-1}'$ and $\hat{s}_j$. To visualize this, we show below the explicit form of $R(2)$, $R(3)$ and $R(4)$.
\begin{align}
R(2) &= \begin{bmatrix}\cos\theta_2 & -\sin\theta_2 & 0 & 0 & \ldots & 0 \\
\sin\theta_2 & \cos\theta_2 & 0 & 0 & \ldots & 0 \\
0 & 0 & 1 & 0 & \ldots & 0 \\
0 & 0 & 0 & 1 & \ldots & 0 \\
\vdots & \vdots & \vdots & \vdots & \ddots & \vdots \\
0 & 0 & 0 & 0 & \ldots & 1\end{bmatrix},\nonumber\\
\theta_2 &= \cos^{-1}(1/\sqrt{2}),\nonumber\\
R(3) &= \begin{bmatrix}1 & 0 & 0 & 0 & \ldots & 0 \\
0 & \cos\theta_3 & -\sin\theta_3 & 0 & \ldots & 0 \\
0 & \sin\theta_3 & \cos\theta_3 & 0 & \ldots & 0 \\
0 & 0 & 0 & 1 & \ldots & 0 \\
\vdots & \vdots & \vdots & \vdots & \ddots & \vdots \\
0 & 0 & 0 & 0 & \ldots & 1\end{bmatrix},\nonumber\\
\theta_3 &= \cos^{-1}(1/\sqrt{3}),\nonumber\\
R(4) &= \begin{bmatrix}1 & 0 & 0 & 0 & \ldots & 0 \\
0 & 1 & 0 & 0 & \ldots & 0 \\
0 & 0 & \cos\theta_4 & -\sin\theta_4 & \ldots & 0 \\
0 & 0 & \sin\theta_4 & \cos\theta_4 & \ldots & 0 \\
\vdots & \vdots & \vdots & \vdots & \ddots & \vdots \\
0 & 0 & 0 & 0 & \ldots & 1\end{bmatrix},\nonumber\\
\theta_4 &= \cos^{-1}(1/\sqrt{4}).
\end{align}

In general, for arbitrary $N$, $n$ and therefore, $J$, the SCS and ACS's can be expressed as follows
\begin{align}
\ket{E_n'}=&\sum_{l=1}^J \hat{s}_l/\sqrt{J}, \nonumber\\
\ket{E_{n,j}'}=&(\sum_{l=1}^j \hat{s}_l-j\hat{s}_{j+1})/\sqrt{j(j+1)},
\label{Eq_Collective_States_Definition}
\end{align}
where $j=1, 2, \ldots, n-1$. Conversely, the original unrotated vectors can be written in terms of the rotated, collective states bases as follows
\begin{align}
\hat{s}_1 =& \ket{E_n'}/\sqrt{J}+\sum_{j=1}^{J-1} \ket{E_{n,j}'}/\sqrt{j(j+1)}, \nonumber\\
\hat{s}_j =& \ket{E_n'}/\sqrt{J}+\sum_{l=j}^{J-1}\ket{E_{n,l}'}/\sqrt{l(l+1)}-\sqrt{j-1}\ket{E_{n,j-1}'}/\sqrt{j},
\label{Eq_Collective_States_Definition_Inverse}
\end{align}
where $j=2, \ldots, n-1$. This inversion is useful in illustrating the behavior of the collective states in more complex situations, an example of which will be presented shortly.

In order to get a clearer picture of how the spread in detuning affects the behavior of the ensemble, we consider the simple case of a 3-atom ensemble interacting with a laser of uniform profile. Additionally, we assume that the $i$-th atom experiences a detuning of $\delta_i$. The manifold corresponding to the absorption of $1$ photon is spanned by the set $\Phi_{3,1}$, whose elements, given by $\ket{e_1'g_2'g_3'},\ket{g_1'e_2'g_3'}$ and $\ket{g_1'g_2'e_3'}$, are now labeled as $\hat{s}_1,\hat{s}_2$ and $\hat{s}_3$, respectively. The SCS of this manifold, as defined in Eq.~(\ref{Eq_Collective_States_Definition}), is given by $\ket{E_1'}=(\hat{s}_1+\hat{s}_2+\hat{s}_3)/\sqrt{3}=(\ket{e_1'g_2'g_3'}+\ket{g_1'e_2'g_3'}+\ket{g_1'g_2'e_3'})/\sqrt{3}$. One of the possible ways of forming the set of ACS's is $\ket{E_{1,1}'}=(\hat{s}_1-\hat{s}_2)/\sqrt{2}=(\ket{e_1'g_2'g_3'}-\ket{g_1'e_2'g_3'})/\sqrt{2}$, and $\ket{E_{1,2}'}(\hat{s}_1+\hat{s}_2-2\hat{s}_3)/\sqrt{6}=(\ket{e_1'g_2'g_3'}+\ket{g_1'e_2'g_3'}-2\ket{g_1'g_2'e_3'})/\sqrt{6}$. The action of the ensemble Hamiltonian, $H_C'=H_r'+H_l'+H_d'$ on $\ket{E_1'}$ shows how it experiences an energy shift, and couples with its adjacent states as follows:
\begin{subequations}
\begin{align}
H_r'\ket{E_1'}/\hbar &= \Omega(\ket{e_1'e_2'g_3'}+\ket{e_1'g_2'e_3'}+\ket{g_1'e_2'e_3'})/\sqrt{3},
\label{Eq_3atom_Hr}
\end{align}
\begin{align}
H_l'\ket{E_1'}/\hbar &= \sqrt{3}\Omega\ket{g_1'g_2'g_3'}/2,
\label{Eq_3atom_Hl}
\end{align}
\begin{align}
H_d'\ket{E_1'}/\hbar &= (-\delta_1\ket{e_1'g_2'g_3'}-\delta_2\ket{g_1'e_2'g_3'}-\delta_3\ket{g_1'g_2'e_3'})/\sqrt{3}.
\label{Eq_3atom_Hd}
\end{align}
\end{subequations}
It can be seen from Eq.~(\ref{Eq_Collective_States_Definition}) that Eq.~(\ref{Eq_3atom_Hr}) can be written as $H_r'\ket{E_1'}/\hbar = \Omega\ket{E_2'}$ and Eq.~(\ref{Eq_3atom_Hl}) can be written as $H_l'\ket{E_1'}/\hbar = \sqrt{3}\Omega\ket{G'}/2$. Furthermore, each term on the right hand side in Eq.~(\ref{Eq_3atom_Hd}) can be written in terms of the relevant SCS and ACS's according to Eq.~(\ref{Eq_Collective_States_Definition_Inverse}):
\begin{align}
H_d'\ket{E_1'}/\hbar =& -\delta_1\hat{s}_1/\sqrt{3}-\delta_2\hat{s}_2/\sqrt{3}-\delta_3\hat{s}_3/\sqrt{3} \nonumber\\
=& -(\delta_1+\delta_2+\delta_3)\ket{E_1'}/3-(\delta_1-\delta_2)\ket{E_{1,1}'}/\sqrt{6}\nonumber\\
&-(\delta_1+\delta_2-2\delta_3)\ket{E_{1,2}'}/\sqrt{18}.
\label{Eq_3atom_E1_evolution_Collective}
\end{align}
The first term in parentheses on the right hand side of Eq.~(\ref{Eq_3atom_E1_evolution_Collective}) is the energy shift in $\ket{E_1'}$. The second and third terms give the coupling strength of $\ket{E_1'}$ with $\ket{E_{1,1}'}$ and $\ket{E_{1,2}'}$, respectively. In the case that each atom in the ensemble experiences the same detuning due to Doppler shift, these two terms go to zero, and the ACS's remain uncoupled from the symmetric set. 

In the more complex case where each atom in the ensemble experiences a unique Rabi frequency, the raising part of the ensemble Hamiltonian applied to any SCS yields the next higher SCS of that ensemble, as prescribed in Eq.~(\ref{Eq_En+1_from_En}). To illustrate this, we consider the example of a $4$-atom ensemble where the raising part of the Hamiltonian is $H_{r}'={\sum}_{i=1}^4\hbar\Omega_i\ket{e_i'}\bra{g_i'}/2$. The set of SCS's are therefore, the following:
\begin{align}
\ket{E_1'}=&(\Omega_1\ket{e_1'g_2'g_3'g_4'}+\Omega_2\ket{g_1'e_2'g_3'g_4'}+\Omega_3\ket{g_1'g_2'e_3'g_4'}+\Omega_4\ket{g_1'g_2'g_3'e_4'})\nonumber\\
&\times(\Omega_1^2+\Omega_2^2+\Omega_3^2+\Omega_4^2)^{-1/2}\nonumber\\
\ket{E_2'}=&(\Omega_1\Omega_2\ket{e_1'e_2'g_3'g_4'}+\Omega_1\Omega_3\ket{e_1'g_2'e_3'g_4'}+\Omega_1\Omega_4\ket{e_1'g_2'g_3'e_4'}+\Omega_2\Omega_3\ket{g_1'e_2'e_3'g_4'}\nonumber\\
&+\Omega_2\Omega_4\ket{g_1'e_2'g_3'e_4'}+\Omega_3\Omega_4\ket{g_1'g_2'e_3'e_4'})((\Omega_1\Omega_2)^2+(\Omega_1\Omega_3)^2+(\Omega_1\Omega_4)^2\nonumber\\
&+(\Omega_2\Omega_3)^2 +(\Omega_2\Omega_4)^2+(\Omega_3\Omega_4)^2)^{-1/2}\nonumber\\
\ket{E_3'}=&(\Omega_1\Omega_2\Omega_3\ket{e_1'e_2'e_3'g_4'}+\Omega_1\Omega_2\Omega_4\ket{e_1'e_2'g_3'2_4'}
+\Omega_1\Omega_3\Omega_4\ket{e_1'g_2'e_3'e_4'}\nonumber\\
&+\Omega_2\Omega_3\Omega_4\ket{g_1'e_2'e_3'e_4'})((\Omega_1\Omega_2\Omega_3)^2+(\Omega_1\Omega_2\Omega_4)^2+(\Omega_1\Omega_3\Omega_4)^2\nonumber\\
&+(\Omega_2\Omega_3\Omega_4)^2)^{-1/2}\nonumber\\
\ket{E_4'}=&\ket{e_1'e_2'e_3'e_4'}.
\label{Eq_Collective_symmetric_states_4atoms}
\end{align}

The set of ACS's corresponding to $\ket{E_n'}$ in the present case of non-uniform Rabi frequency consists of $(J-1)$ elements that are orthogonal to one another as well as to $\ket{E_n'}$. As mentioned above, they can be constructed using the GSO process. The realization of this process as a set of rotations follows a similar set of rules as described above. However, the rotation angles will now depend on the relative amplitudes of all the Rabi frequencies. The details of this process are beyond the scope of the present discussion and will be presented elsewhere.

\section{Quantized COM Model of Ensemble}
\label{sec:QuantizedCOM}

As mentioned earlier, we have been investigating the use of atomic ensembles for a Collective State Atomic Interferometer (COSAIN). In a Conventional Raman Atomic Interferometer (CRAIN), one must take into account the quantum nature of the COM motion. Similarly, for a COSAIN, we must consider the COM motion of the atom quantum mechanically. In doing so, one must consider all the degrees of freedom of the COM. However, for a CRAIN as well as the COSAIN (which is a variant of the CRAIN), only the motion in the direction parallel to the laser beams (which we have chosen to be the \textbf{z} direction) has to be quantized. As such, in what follows, we keep our discussion confined to such a scenario.

The $i$-th atom is now a Gaussian wavepacket formed by the superposition of an infinite number of plane waves, where the $p$-th plane wave can exist in two energy states, 
$\ket{g_{ip},\hbar{k_{ip}'}}$ and $\ket{e_{ip},\hbar(k_{ip}'+k)}$, which differ by a momentum $\hbar k$. Since the laser field amplitude is assumed to be constant in the \textbf{z} direction, the Rabi frequency experienced by each plane wave manifold of the $i$-th atom is $\Omega_i$. The Doppler shift induced due to the thermal motion of the atoms in the \textbf{z} direction ascribes a detuning of $\delta_{Ti}$ to this atom. As such, the Hamiltonian of the $p$-th plane wave of the $i$-th atom is
\begin{align}
H_{ip}/\hbar =& (\omega_g+\hbar k_{ip}'^2/2m)\ket{g_{ip}}\bra{g_{ip}}\nonumber\\
&+(\omega_e+\hbar (k_{ip}'+k)^2/2m)\ket{e_{ip}}\bra{e_{ip}} \nonumber\\
&+\Omega_i(\exp{(i(\omega_{0i}t-\xi_i))}\ket{g_{ip}}\bra{e_{ip}}+ h.c.)/2.
\end{align}

\begin{figure}[h]
\centering
\includegraphics[scale=0.2]{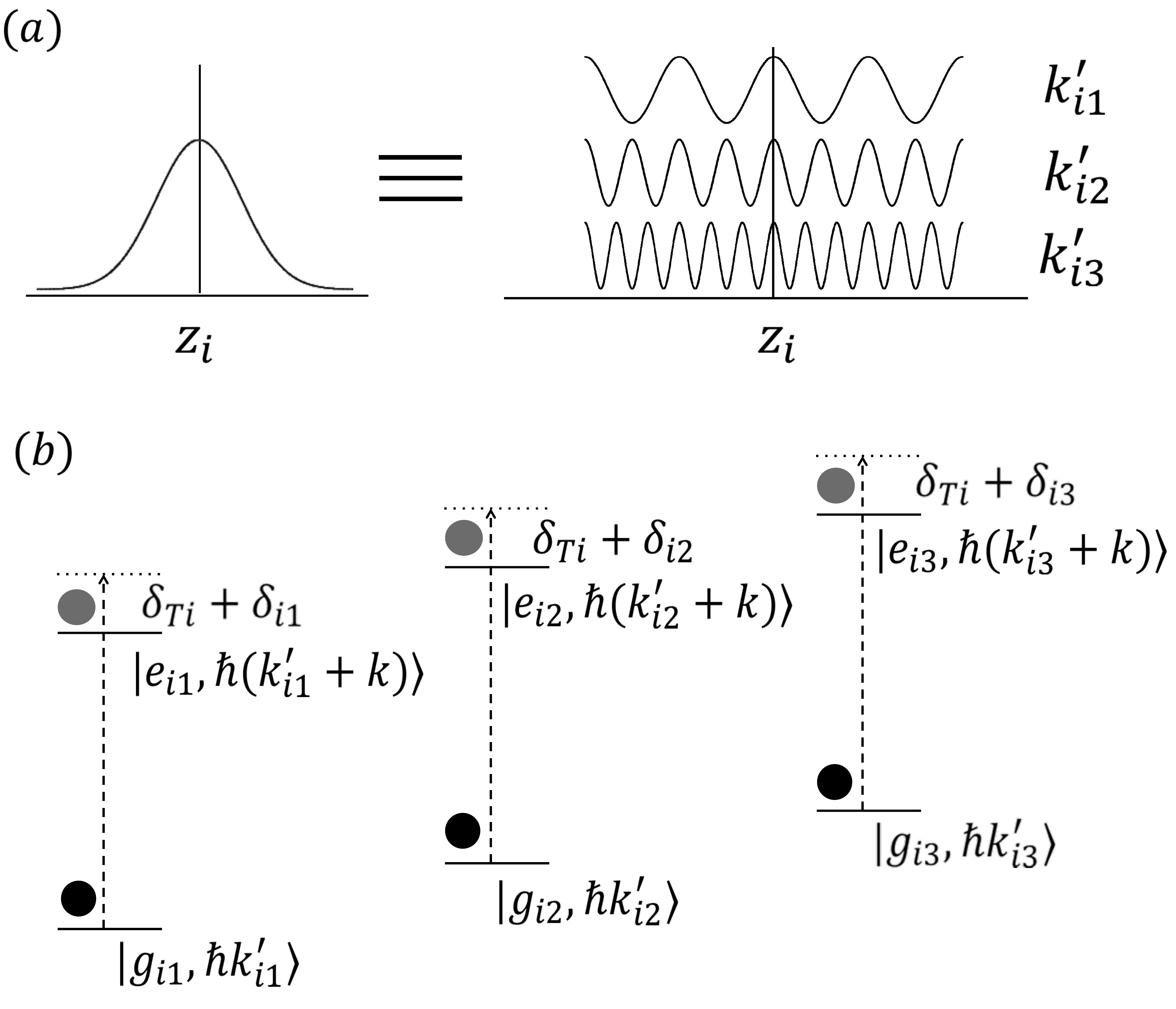}
\caption{\label{Fig_5} (a) Quantized COM model of an atom, (b) two level model of each plane wave component.}
\end{figure}

The Schr\"odinger equation governing the evolution of the state vector of this plane wave, $\ket{\psi_{ip}}$, is $i\hbar\partial\ket{\psi_{ip}}/\partial{t}=H_{ip}\ket{\psi_{ip}}$. Similar to the description given in Eq.~(\ref{eq_SingleQ}-\ref{eq_SingleHam_Qtransformed}), a unitary transformation, $Q_{ip}$ changes $\ket{\psi_{ip}}$ to $\ket{\psi_{ip}'}$ such that
\begin{align}
Q_{ip} = \sum_{j=1}^2\exp{(i(a_{ipj}t+b_{ipj}))}\ket{j}\bra{j}, 
\end{align}
where $a_{ipj}$ and $b_{ipj}$ are arbitrary parameters. The Hamiltonian in the new basis vector thus formed is $H_{ip}'=Q_{ip}H_{ip}Q_{ip}^{-1}-\hbar\dot{Q}_{ip}Q_{ip}^{-1}$, so that $i\hbar\partial\ket{\psi_{ip}'}/\partial{t}=H_{ip}'\ket{\psi_{ip}'}$. It can be stripped of its time dependence by setting $a_{ip1}=\omega_g+\hbar k_{ip}'^2/2m$ and $a_{ip2}=\omega_e+\delta_{vi}+\hbar k_{ip}'^2/2m$. For $b_{ip1}=0$ and $b_{ip2}=-\xi_i$, $H_{ip}'$ is rendered independent of any phase factors. In the transformed frame, the Hamiltonian is thus
\begin{align}
H_{ip}'/\hbar =& (-\delta_{vi}+\hbar k^2/2m+\hbar kk'_{ip}/m)\ket{e_{ip}'}\bra{e_{ip}'}+\Omega_i(\ket{g_{ip}'}\bra{e_{ip}'}+h.c.)/2.
\end{align}

Since the atom is a sum of these individual plane waves, it evolves according to the equation that is the sum of the individual Schr\"odinger equations, $i\hbar\partial({\sum}_{p\rightarrow-\infty}^{\infty}\ket{\psi_{ip}'})/\partial{t}={\sum}_{p\rightarrow-\infty}^{\infty}H_{ip}'\ket{\psi_{ip}'}$. In the limit that the Rabi frequency of the $i$-th atom is large compared to the Doppler shift due to the COM momentum of each of the constituent plane waves, i.e. $\Omega_i\gg\hbar kk'_{ip}/m$ , the corresponding Hamiltonians become identical to one another. The resulting evolution equation is then $i\hbar\partial\ket{\psi_{i}'}/\partial{t}=H_{i}'\ket{\psi_{i}'}$, where $\ket{\psi_{i}'}=\sum_{p\rightarrow-\infty}^{\infty}\ket{\psi_{ip}'}$ and $H_i'=H_{i1}'=H_{i2}'$, etc. In this regime, the atom's Hamiltonian becomes $H_{i}'/\hbar=-\delta_i\ket{e_{i}'}\bra{e_{i}'}+\Omega_i(\ket{g_{i}'}\bra{e_{i}'}+h.c.)/2$, where $\delta_i=\delta_{vi}-\hbar k^2/2m$. This is identical to the semiclassical Hamiltonian of the atom where the COM mass degree of freedom of the atom is not considered. Thus, we conclude that, under approximations that are valid for the COSAIN, a semi-classical description of the COM motion of each atom is sufficient. As such, all the results we have derived above  regarding the properties of collective state remain valid for the COSAIN.

\section{Summary}
We have investigated the behavior of an ensemble of $N$ non-interacting, identical atoms, excited by a laser with a wavelength of $\lambda$. In doing so, we have assumed that the wavefunctions of the atoms do not overlap with one another, so that quantum statistical properties are not relevant. In general, the $i$-th atom sees a Rabi frequency $\Omega_i$, an initial position dependent laser phase $\phi_i$, and a motion induced Doppler shift of $\delta_i$. When $\Omega_i = \Omega$ and $\delta_i = \delta$ for all atoms, the system evolves into a superposition of $(N+1)$ generalized symmetric collective states, independent of the values of $\phi_i$. If $\phi_i = \phi$ for all atoms, these states simplify to the well known Dicke collective states. When $\Omega_i$ or $\delta_i$ is distinct for each atom, the system evolves into a superposition of symmetric as well as asymmetric collective states. For large values of $N$, the number of asymmetric states $(2^N-(N+1))$ is far larger than that of the symmetric states. For a COSAIN and a COSAC it is important to understand the behavior of all the collective states under various conditions. In this paper, we have shown how to formulate systematically the properties of all the collective states under various non-idealities, and used this formulation to understand the dynamics thereof. Specifically, for the case where $\Omega_i = \Omega$ and $\delta_i = \delta$ for all atoms, we have shown how the amplitudes of each of the generalized collective states can be determined explicitly in a simple manner. For the case where $\Omega_i$ or $\delta_i$ is distinct for each atom, we have shown how the symmetric and asymmetric collective states can be treated on the same footing. Furthermore, we have shown that the collective states corresponding to the absorption of a given number of photons can be visualized as an abstract, multi-dimensional rotation in the Hilbert space spanned by the ordered product states of individual atoms. This technique enables one to construct the explicit expression for any asymmetric state of interest. Such expressions in turn can be used to determine the evolution of such a state in the COSAIN or the COSAC. We have also considered the effect of treating the COM degree of freedom of the atoms quantum mechanically on the description of the collective states. This is particularly relevant for the COSAIN. In particular, we have shown that it is indeed possible to construct a generalized collective state when each atom is assumed to be in a localized wave packet.

\appendix
\counterwithin{figure}{section}
\section{\label{sec:append}Equivalence Between Doppler Effect Induced Phase Shift and Position Change Induced Phase Shift}
\begin{figure}[h]
\centering
\includegraphics[scale=0.15]{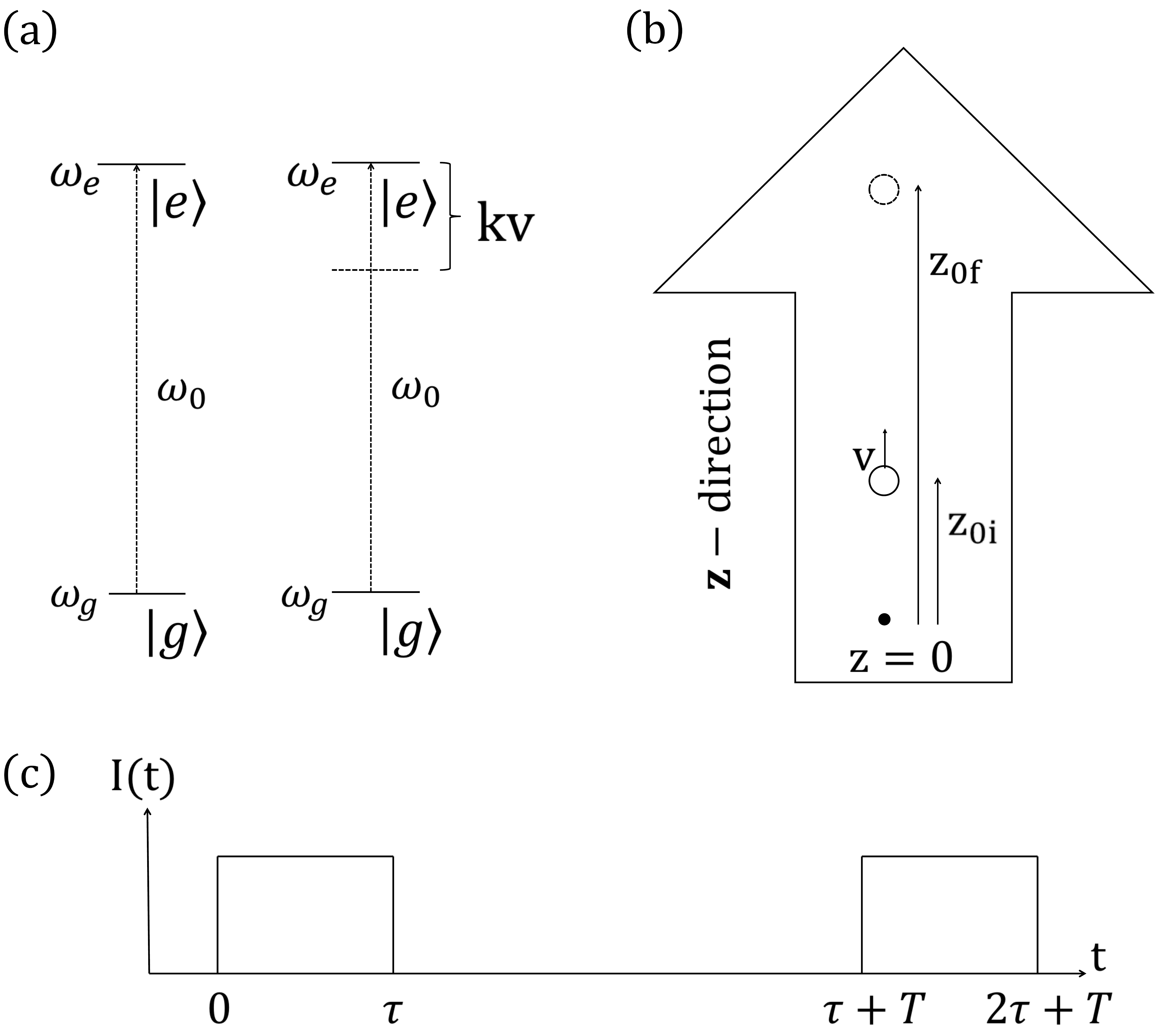}
\caption{\label{Fig_A1} (a) (left) Two level atom in the lab frame frame, (right) in the atom's frame of reference, (b) change in the coordinates of the atom over the duration of interaction with the laser pulses, (c) laser beam intensity variation over the duration of interaction.}
\end{figure}

Consider an ideal two level atom, excited by a laser field traveling in the $\mathbf{z}$ direction, assuming the field amplitude to be uniform in all directions. The atom is modeled as having two energy levels, $\ket{g}$ and $\ket{e}$. For the issue at hand, it is not necessary to consider the radiative decay of $\ket{e}$. As such, we assume both of the states to be long-lived. This two-level system is shown in Fig.~\ref{Fig_A1}(a), where $\omega_0$ is the frequency of the laser field, assumed to be resonant for a stationary atom. The laser field is assumed to be polarized, arbitrarily, in the $\mathbf{x}$ direction. As illustrated in Fig.~\ref{Fig_A1}(b), the atom is initially $(t=0)$ positioned at $\mathbf{r_{0i}}=x_{0i}\mathbf{\hat{x}}+y_{0i}\mathbf{\hat{y}}+z_{0i}\mathbf{\hat{z}}$ and is moving in the $\mathbf{z}$ direction with a non-relativistic velocity $v$. The electric field at a time $t$, in the atom's frame of reference, is $\mathbf{E}(\mathbf{r},t)=\mathbf{\hat{x}}E_0\cos(\omega_0t-kz)$, where $z=z_{0i}+vt$. In the semiclassical model employed here, the Hamiltonian of this atom can be written as $H=H_{0i}+q\mathbf{\rho}\cdot\mathbf{E}$, where the terms have their usual meanings as given in Sec.~\ref{sec:semiclass}. After making the rotating-wave approximation as prescribed in Sec.~\ref{sec:semiclass}, $H$ can be written in the bases of $\ket{g}$  and $\ket{e}$ as
\begin{align}
H/\hbar =& \omega_g\ket{g}\bra{g}+\omega_e\ket{e}\bra{e}+\Omega(\exp(i(\omega_0t-k(z_{0i}+vt)))\ket{g}\bra{e}+h.c.)/2,
 \label{eq_SingleHam_appendix}
\end{align}
where $\Omega\equiv\bra{g}(\mathbf{x}\cdot\mathbf{\rho})\ket{e}E/\hbar=\bra{e}(\mathbf{x}\cdot\mathbf{\rho})\ket{g}E/\hbar$.

The state of this atom, $\ket{\psi}$ evolves according to the Schr\"odinger equation, $i\hbar\partial\ket{\psi}/\partial{t}=H\ket{\psi}$. A unitary transformation, $Q$ defined as $Q= \sum_{j=1}^{2}\exp(i(a_{j}t+b_{j}))\ket{j}\bra{j}$ changes $\ket{\psi}$ to $\ket{\psi'}=Q\ket{\psi}$, where $a_{j}$ and $b_{j}$ are arbitrary parameters. The Q-transformed Hamiltonian for this state vector is then $H'=QHQ^{-1}-\hbar\dot{Q}Q^{-1}$, so that $i\hbar\partial\ket{\psi'}/\partial{t}=H'\ket{\psi'}$. $H'$ is stripped of its time dependence by setting $a_{1}=\omega_g$ and $a_{2}=\omega_{0}+\omega_g=\omega_e-kv$. Now, setting $b_{1} = 0$, $b_{2} = -kz_{0i}$ makes $H'$ independent of any phase factor as well. The $Q$-transformed Hamiltonian thus becomes
\begin{align}
 H'/\hbar = kv\ket{e'}\bra{e'}+\Omega(\ket{g'}\bra{e'}+ h.c.)/2.
      \label{eq_SingleHam_Qtransformed_appendix}
\end{align}
Therefore, the velocity of the atom induces a net detuning of $\delta =-kv$. The new basis vectors, $\ket{g'}$ and  $\ket{e'}$, are related to the original basis vectors as $\exp(-i\omega_gt)\ket{g}$ and $\exp(-i((\omega_e-kv)t-kz_{0i}))\ket{e}$, respectively. If the atom is initially in $c_{gi}(0)\ket{g_i'}+c_{ei}(0)\ket{e_i'}$, its state after interaction for a time $t$ is given by Eq.~(\ref{eq_SA_evolution_general}).

We consider this atom's interaction with two consecutive laser fields separated by a dark zone of duration $T$, as illustrated in Fig.~\ref{Fig_A1}(c). The time of interaction of the atom with each field is such that $\tau = \pi/2\Omega$. The atom initially at $z=z_{0i}$ drifts to $z=z_{0f}$ by the end of the entire interaction sequence. For the sake of simplicity, we assume that $kv \ll \Omega$ and that the atom's position does not change appreciably over the duration of the pulse. Starting with the atom in state $\ket{g}$ at $t=0$, the state of the atom at the end of the first pulse $(t = \tau)$ is $\ket{\psi'}=(\ket{g'}-i\ket{e'})/\sqrt(2)$. The $Q$-transformed Hamiltonian in the dark zone is given by $H_d'=kv\ket{e'}\bra{e'}$. At $t =\tau+T$, the state of the atom can be expressed as 
\begin{equation}
\ket{\psi'}=(\ket{g'}-i\exp(-ikvT)\ket{e'})/\sqrt{2}.
\label{Eq_end_Dark_zone}
\end{equation}
After the atom's encounter with the second pulse $(t=2\tau+T)$, its quantum state can be written as $\ket{\psi'}=(1-\exp(-ikvT))\ket{g'}/2-i(1+\exp(-ikvT))\ket{e'}/2$. In the original bases of $\ket{g}$ and $\ket{e}$, the final state of the atom at the end of the separated field interaction sequence is given by
\begin{align}
\ket{\psi}=&(1-\exp(-ikvT))\exp(-i\omega_gt)\ket{g}/2-i(1+\exp(-ikvT))\nonumber\\
&\times \exp(-i(\omega_e-kv)t+ikz_{0i})\ket{e}/2.
\label{Eq_atoms_frame_final}
\end{align}

Now, we consider the same interaction shown in Fig.~\ref{Fig_A1}(c) in the laboratory frame of reference in which the electric field at any point along the laser's direction of propagation (\textbf{z} direction) is given by $\mathbf{E}(\mathbf{r},t)=\mathbf{\hat{x}}E_0\cos(\omega_0t-kz)$. Considering that at $t=0$ the atom is positioned at $z=z_{0i}$, the Hamiltonian for the first interaction zone is given by $H_{L1}/\hbar = \omega_g\ket{g}_{LL}\!\bra{g}+\omega_e\ket{e}_{LL}\!\bra{e}+\Omega(\exp(i(\omega_0t-kz_{0i}))\ket{g}_{LL}\!\bra{e}+h.c.)/2$, where the subscript $L$ indicates that this is in the laboratory frame. The state of the atom evolves according to $i\hbar\partial\ket{\psi}_L/\partial{t}=H_{L1}\ket{\psi}_L$. Therefore, the transformation $Q_1$ to remove time and phase dependence from $H_{L1}$ is given by $Q_1=\exp(i\omega_gt)\ket{1}\bra{1}+\exp(i(\omega_et-kz_{0i}))\ket{2}\bra{2}$. The resulting $Q$-transformed Hamiltonian in the bases of $\ket{g'}_L$ and $\ket{e'}_L$ is $H_{L1}'/\hbar=(\Omega\ket{g'}_{LL}\!\bra{e'}+h.c.)/2$. As a result, considering that the atom is in state $\ket{g'}_L$ at $t=0$, the state of the atom at $t=\tau$ is $\ket{\psi'}=(\ket{g'}_L-i\ket{e'}_L)/\sqrt{2}$.

The dark zone $Q$-transformed Hamiltonian, $H_{Ld}'$ contains no non-zero elements. Thus, at the end of the dark zone $(t=\tau+T)$, the quantum state of the atom remains unaltered. Since the atom has a non zero velocity, $v$ along \textbf{z} direction, by the end of the dark zone it will have moved to $z=z_{0f}$. As a consequence, the Hamiltonian for the second pulse will be $H_{L2}/\hbar = \omega_g\ket{g}_{LL}\!\bra{g}+\omega_e\ket{e}_{LL}\!\bra{e}+\Omega(\exp(i(\omega_0t-kz_{0f}))\ket{g}_{LL}\!\bra{e}+h.c.)/2$. The $Q$-transformation required to make $H_{L2}$ time and phase factor independent may be written as $Q_2 = \exp(i\omega_gt)\ket{1}\bra{1}+\exp(i(\omega_et-kz_{0f}))\ket{2}\bra{2}$ and we define $\ket{\psi''}_L=Q_2\ket{\psi}_L$. The new basis states thus formed are $\ket{g''}_L=\exp(i\omega_gt)\ket{g}_L$ and $\ket{e''}_L=\exp(i(\omega_et-kz_{0f}))\ket{e}_L$. Therefore, the quantum state of the atom at the end of the dark zone $(t=\tau+T)$, must now be written in the $Q_2$-transformed bases of $\ket{g''}_L$ and $\ket{e''}_L$. As such, we get $\ket{\psi''}_L=Q_2Q_1^{-1}\ket{\psi'}_L=(\ket{g''}_L-i\exp(ik(z_{0i}-z_{0f}))\ket{e''}_L)/\sqrt{2}$. This is the initial condition for the second pulse. At the end of the second pulse, $t=2\tau+T$, the atom's quantum state is, therefore, given by $\ket{\psi''}_L=(1-\exp(ik(z_{0i}-z_{0f})))\ket{g''}_L/2-i(1+\exp(ik(z_{0i}-z_{0f})))\ket{e''}_L/2$. Thus, in the original bases of $\ket{g}_L$ and $\ket{e}_L$, the state of the atom is 
\begin{align}
\ket{\psi}_L=&(1-\exp(ik(z_{0i}-z_{0f})))\exp(-i\omega_gt)\ket{g}_L/2\nonumber\\
&-i(1+\exp(ik(z_{0i}-z_{0f})))\exp(-i\omega_et+ikz_{0f})\ket{e}_L/2.
\label{Eq_lab_frame_final}
\end{align}

Since $z_{0f}=z_{0i}+vT$, Eq.~(\ref{Eq_lab_frame_final}) is identical to Eq.~(\ref{Eq_atoms_frame_final}). Thus, when one takes into account the Doppler shift, it is no longer necessary to consider explicitly the fact that the atom sees a different laser phase at different times.

This work has been supported by the NSF grants number DGE-0801685 and DMR-1121262, and AFOSR grant number FA9550-09-1-0652.

\end{document}